\documentclass[aps,twocolumn,PRL]{revtex4}

\usepackage{graphics}
\usepackage{graphicx}
\usepackage{bm}
\usepackage{latexsym}
\usepackage{color, soul}
\usepackage{mathrsfs}
\usepackage{braket}
\usepackage{enumerate}
\usepackage[usenames,dvipsnames]{xcolor}
\usepackage[colorlinks=true,citecolor=blue,linkcolor=blue,urlcolor=blue,backref=true]{hyperref}
\usepackage{float}
\usepackage[raggedright,tight]{subfigure}  
\usepackage{amsmath}
\usepackage{amsfonts}
\usepackage{braket}
\usepackage[normalem]{ulem}
\usepackage{cancel}
\usepackage{mathtools}
\usepackage{appendix}

\begin{document}

\title{Dicke Superradiance in Extended 2D Quantum Arrays Coupled to Metasurface Bound States in the Continuum }

\author{Daniel Eyles$^{1,3}$}
\author{Emmanuel Lassalle$^3$}
\author{Adam Stokes$^2$}
\author{Ahsan Nazir$^1$}
\author{Ramón Paniagua-Domínguez$^{3,4}$}

\affiliation{$^1$Department of Physics and Astronomy, University of Manchester, Oxford Road, Manchester M13 9PL, United Kingdom}
\affiliation{$^2$School of Mathematics, Statistics, and Physics, Newcastle University, Newcastle upon Tyne NE1 7RU, United Kingdom}
\affiliation{$^3$Institute of Materials Research and Engineering, A*STAR (Agency for Science, Technology and Research), 2 Fusionopolis Way, Innovis \#08-03, 138634 Singapore, Republic of Singapore}
\affiliation{$^4$Instituto de Estructura de la Materia (IEM), Consejo Superior de Investigaciones Científicas, Serrano 121, 28006 Madrid, Spain}

\begin{abstract}
Dicke superradiance is a collective phenomenon where the emission from ensembles of quantum emitters is coherently enhanced beyond the sum of each emitter’s independent emission. Here, we propose a platform that exploits the delocalised nature of a high-Q, non-local mode supported by a dielectric metasurface (a so-called bound-state-in-the-continuum or BIC) to induce superradiant behaviour within an extended two-dimensional array of distant quantum emitters. We show that these BIC-mediated emitter interactions can span several wavelengths, thus overcoming the traditional subwavelength separation between emitters required in free space. We further show that reaching the idealised Dicke limit is possible in this system, provided that the emitters are coupled to the BIC mode efficiently enough, as quantified through the $\beta$-factor. Moreover, we demonstrate its experimental viability by analysing its robustness to realistic experimental imperfections. This work puts forward optical metasurfaces supporting BICs as a physically viable platform for realising the upper limits of cooperative emission in physically extended quantum emitter arrays.
\end{abstract}

\maketitle

\newpage

\emph{Introduction}—Collective phenomena play a central role in quantum optics, giving rise to emergent behaviour beyond that of individual quantum systems. A paradigmatic example is \textit{superradiance}, first studied by Dicke~\cite{Dicke_Paper}, in which ensembles of indistinguishable quantum emitters (QEs) radiate cooperatively through shared electromagnetic modes. Synchronized dipole oscillations lead to constructive interference, producing emission intensities that scale as \(N^2\) and a collectively enhanced decay rate proportional to \(N\)~\cite{Dicke_Superradiance_4, Dicke_Superradiance_5}.

Superradiance is also encoded in the photon statistics of the emitted field. The second-order correlation function, \(\mathcal{G}^{(2)}(t,\tau)\), quantifies photon coincidences, with \(\mathcal{G}^{(2)}(t,0)>1\) indicating cooperative photon bunching~\cite{Dicke_Superradiance_3}. In free space, however, superradiance is restricted to emitters separated by distances much smaller than the emission wavelength, \(\lambda_0 = 2\pi c/\omega_0\)~\cite{Dicke_Paper}. Even in ordered atomic or solid-state arrays~\cite{Garcia_Dimensionality, Garcia_Universal, robicheaux2021theoretical, mok2023dicke}, cooperative emission is rapidly suppressed with increasing separation; for example, in a \(20\times20\) square lattice, \(\mathcal{G}^{(2)}(0)>1\) is only observed for inter-emitter spacings \(d \lesssim 0.4\lambda_0\)~\cite{Garcia_Universal}. This severely constrains scalability and experimental control.

Nanophotonic environments offer routes to overcome these limitations by mediating long-range interactions between distant emitters. One-dimensional waveguides enable effectively infinite-range coupling along the propagation direction~\cite{Exp_Waveguide1, asenjo2017exponential, Exp_Waveguide2, solano2017super}, while optical metasurfaces provide scalable control of light–matter interactions in two dimensions~\cite{Chen_2016, Chen2020-gk, Hu2021-kt, Zhao_Review_Meta}. Nevertheless, existing approaches rely on reduced dimensionality, localized resonances, or finite mode volumes, and a mechanism enabling genuinely nonlocal cooperative emission across extended two-dimensional arrays remains unexplored.

Here we identify bound states in the continuum (BICs) as a qualitatively new mediator of collective emission. BICs are symmetry-protected or interference-induced optical modes that, despite residing within the radiation continuum, exhibit diverging quality factors and highly delocalized field profiles~\cite{BIC_Theory1, BIC_Theory2}. In dielectric metasurfaces, quasi-BICs retain strong spatial extension while remaining accessible to emitters, enabling long-range electromagnetic correlations~\cite{riley2025metasurfacemediatedquantumentanglementbound}. Unlike cavity or waveguide modes, BICs combine high coherence with nonlocality in two dimensions, providing a distinct route to cooperative light–matter interactions.

We demonstrate that a dielectric metasurface supporting a delocalized quasi-BIC enables Dicke superradiance in extended two-dimensional emitter arrays, even when inter-emitter separations greatly exceed the emission wavelength. The system approaches the ideal Dicke limit of the second-order correlation function, with the degree of cooperativity governed by the coupling efficiency to the BIC. Importantly, this regime is achieved without subwavelength spacing and remains robust against disorder and incomplete lattices. These results establish BIC metasurfaces as a new platform for scalable superradiant light–matter interactions in spatially extended systems.

\emph{Theory}: To quantify superradiance, we use the normalized second-order correlation function \(\mathcal{G}^{(2)}(t,\tau)\), which characterizes photon independent of emission direction and polarization. This follows the framework established by Canaguier-Durand and Carminati for two emitters \cite{Carminati_Paper}, and its generalization to $N$ emitters \cite{eyles2026samplingmethodssuperradiancelarge}. The same figure of merit is used in, e.g., Ref.~\cite{Garcia_Universal, Garcia_Dimensionality}.

\begin{figure}[t]
\begin{minipage}{\columnwidth}
\begin{center}
\hspace*{-1mm}\includegraphics[width=0.8\textwidth]{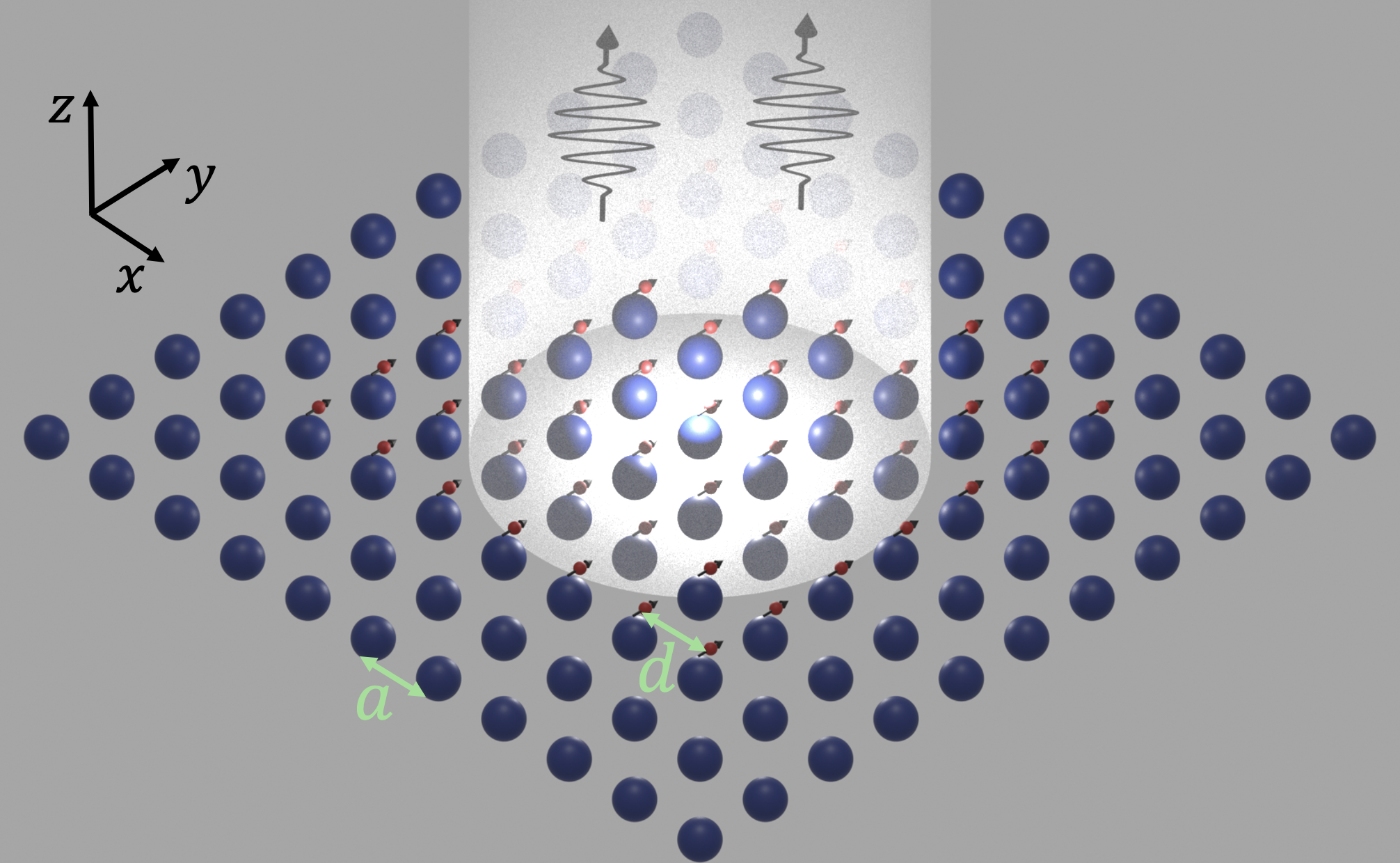}
\caption{A schematic of the QE array (red dots)–metasurface (blue scatterers) system. The emitter positions and dipole moment orientations (black arrows) correspond to the configuration giving maximum coupling to the metasurface-BIC mode. For clarity, only an $11\times 11$ region is shown.}
\label{Lattice Superraidance Diagram}
\end{center}
\end{minipage}
\end{figure}

We consider $N$ identical two-level QEs with transition frequency \(\omega_0\), where \(\ket{\mathrm{g}_\mu}\) and \(\ket{\mathrm{e}_\mu}\) denote the ground and excited states of the $\mu$-th emitter. Each emitter is located at position \(\mathbf{R}_\mu\) and has dipole moment \(\mathbf{d}_\mu\). For a non-absorbing environment, \(\mathcal{G}^{(2)}(t,\tau)\) can be written as \cite{eyles2026samplingmethodssuperradiancelarge}:
\begin{equation}
\begin{split}
&\mathcal{G}^{(2)}(t,\tau) = \\
&\frac{\sum_{\mu,\nu,\gamma,\epsilon} \gamma_{\epsilon\mu}\gamma_{\gamma\nu} 
\langle \hat{\sigma}_\mu^+(t) \hat{\sigma}_\nu^+(t+\tau) 
\hat{\sigma}_\gamma^-(t+\tau) \hat{\sigma}_\epsilon^-(t)\rangle}
{\sum_{\mu,\nu,\gamma,\epsilon}\gamma_{\nu\mu}\gamma_{\epsilon\gamma}
\langle \hat{\sigma}_\mu^+(t+\tau)\hat{\sigma}_\nu^-(t+\tau)\rangle
\langle \hat{\sigma}_\gamma^+(t)\hat{\sigma}_\epsilon^-(t)\rangle},
\end{split}
\label{G2 t}
\end{equation}
where $\hat{\sigma}^-_\mu(t)$ and $\hat{\sigma}^+_\mu(t)$ are the time-dependent QE lowering and raising operators in the Heisenberg picture. The rate $\gamma_{\mu\nu}$ equals the single-emitter rate when $\mu=\nu$ and the dissipative cross rate between the $\mu$-th and $\nu$-th emitters when $\mu\neq\nu$.
It is determined by the dyadic Green's function $\mathbf{G}$ \cite{DGF1, DGF2, DGF3}:
\begin{equation}
\gamma_{\mu\nu} = \frac{2\mu_0 \omega_0^2}{\hbar} 
\mathrm{Im}\big[\mathbf{d}^*_\mu\cdot\mathbf{G}(\mathbf{R}_\mu,\mathbf{R}_\nu,\omega_0)\cdot \mathbf{d}_\nu\big].
\label{rate}
\end{equation}
The spontaneous rate, $\gamma_{\mu\mu}$, are directly related to the local density of states (LDOS), and $\gamma_{\mu\nu}$ ($\mu\neq\nu$) to the cross density of states (CDOS) \cite{carminati2022purcell}, which encodes the electromagnetic coupling between emitters and underpins cooperative emission.


The general expressions above apply to any QE state. In this work, we focus on the case where all emitters are initially excited,
\(\ket{\psi} = \ket{e_\mu}^{\otimes N} = \ket{e_1}\otimes \ket{e_2}\otimes \dots \otimes \ket{e_N}\), i.e. a \textit{fully inverted array} \cite{Fully_Inverted_Array_1, Fully_Inverted_Array_2}. We elect to study this state owing to its experimental accessibility via pulsed excitation, and its fundamental role within cooperative emission, representing the configuration of maximal collective excitation where all emitters radiate coherently, as originally considered by Dicke~\cite{Dicke_Paper}.

For synchronized detection (\(t=\tau=0\)), the expectation values simplify to
\(\langle \sigma_\mu^+ \sigma_\nu^- \rangle = \delta_{\mu\nu}\) and
\(\langle \sigma_\mu^+ \sigma_\nu^+ \sigma_\gamma^- \sigma_\epsilon^- \rangle
= \delta_{\mu\gamma}\delta_{\nu\epsilon} + \delta_{\nu\gamma}\delta_{\mu\epsilon}\),
enabling Eq.~(\ref{G2 t}) to reduce to
\begin{equation}
\begin{split}
\mathcal{G}^{(2)}(0, 0) =  
\frac{\sum_{\mu,\nu,\gamma,\epsilon} \gamma_{\epsilon\mu} \gamma_{\gamma\nu} 
(1-\delta_{\mu\nu})(\delta_{\mu\epsilon}\delta_{\gamma\nu} + \delta_{\mu\gamma}\delta_{\nu\epsilon})}
{\big(\sum_{\mu\nu} \gamma_{\mu\nu} \delta_{\mu\nu}\big)^2 }.
\end{split}
\label{G2 0}
\end{equation}
This removes the explicit dependence on Hilbert-space dimension, eliminating the exponential scaling with $N$ that typically limits numerical simulations \cite{Gardiner_Thermal, eyles2026samplingmethodssuperradiancelarge}. Direct evaluation of the nested summations becomes expensive for large arrays, we generalise the formalism of Ref.~\cite{Garcia_Universal} allowing $\mathcal{G}^{(2)}(0,0)$ to be expressed in terms of eigendecomposition of the decay-matrix, enabling efficient calculation (for the explicit derivation see Supp. Info. Sec.~\ref{Garcia Collective Equation}),
\begin{equation}
\small
    \mathcal{G}^{(2)}(0,0) = 1 + \frac{\sum_\mu^N \Gamma_\mu^2}{\big(\sum_\mu^N \Gamma_\mu\big)^2} - \frac{2 \sum_i^N \Big[ \big( \sum_\mu^N \Gamma_\mu |\alpha_{\mu,i}|^2 \big)^2 \Big]}{\big(\sum_\mu^N \Gamma_\mu\big)^2},
    \label{G2 Garcia Generalised}
\end{equation}
where $\Gamma_\mu$ and $\alpha_\mu$ are the eigenvalues and corresponding eigenvectors of the decay matrix $\bm{\Gamma}|_{\mu\nu} = \gamma_{\mu\nu}$.

\begin{figure*}[t]
\centering
\includegraphics[width=\textwidth]{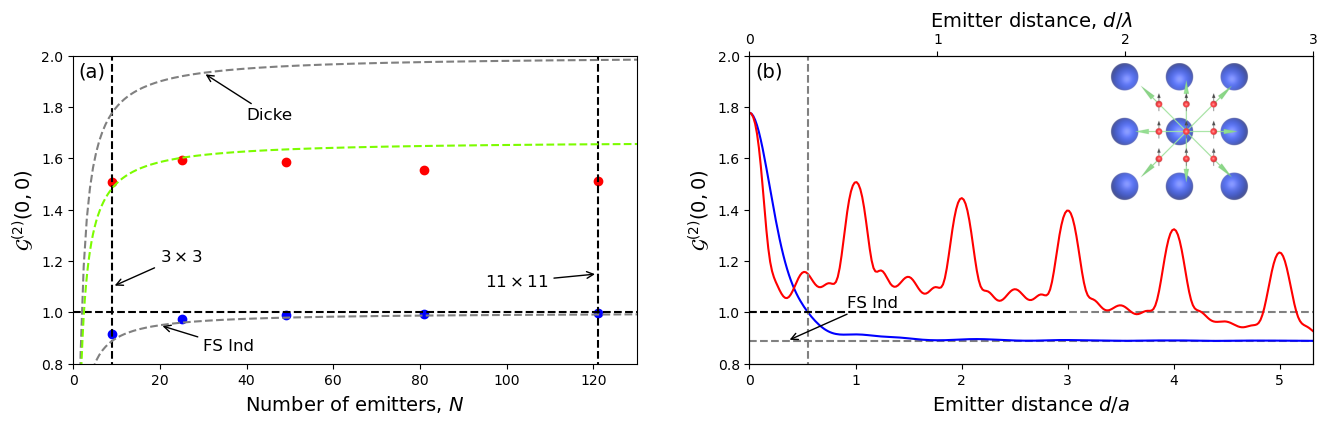}
\caption{\textbf{(a)} $\mathcal{G}^{(2)}(0,0)$ vs $N$ for a finite metasurface (red dots) and free space (blue dots). Dashed green: analytical single-mode model ($\beta=81.79\%$ \cite{riley2025metasurfacemediatedquantumentanglementbound}). Grey: Dicke [Eq.~(\ref{eq:Dicke_result})] and independent [Eq.~(\ref{G2 ind Gen})] limits. \textbf{(b)} $\mathcal{G}^{(2)}(0,0)$ vs lattice constant $d$ for metasurface (red curve) and free space (blue curve). Inset: schematic showing scanning direction for $N=3\times 3$ emitters. Horizontal line indicates the superradiant transition $\mathcal{G}^{(2)}(0,0)=1$.}
\label{Lattice Superraidance}
\end{figure*}


The value of $\mathcal{G}^{(2)}(0,0)$ is bounded between two limiting cases, corresponding to minimal and maximal coupling between emitters. 
In the absence of mutual coupling, i.e.\ for \textit{independent emitters} ($\gamma_{\mu\nu}=0$ for $\mu \neq \nu$), the second-order correlation function reads
\begin{equation}
\mathcal{G}^{(2)}_{\text{Ind}}(0,0) =
1 - \frac{\sum_\mu \gamma_{\mu\mu}^2}{\big(\sum_\mu \gamma_{\mu\mu}\big)^2},
\label{G2 ind Gen}\\    
\end{equation}
where  corresponds to the special case of identical LDOS for all emitters, $\mathcal{G}^{(2)}_{\text{Ind}}(0,0) =
\frac{N-1}{N}$ as would be the case in a homogeneous environment.

At the opposite extreme, for \textit{maximal collective coupling}, $\gamma_{\mu\nu} = \sqrt{\gamma_{\mu\mu}\gamma_{\nu\nu}}$, one obtains
\begin{equation}
\mathcal{G}^{(2)}_{\text{Dicke}}(0,0) =
2 - \frac{2\sum_\mu \gamma_{\mu\mu}^2}{\big(\sum_\mu \gamma_{\mu\mu}\big)^2} = 2 \mathcal{G}^{(2)}_{\text{Ind}}(0,0),
\label{eq:Dicke_result}\\
\end{equation}
which recovers Dicke’s seminal result $\mathcal{G}^{(2)}_{\text{Dicke,H}}(0,0) =
\frac{2(N-1)}{N}$ for identical LDOS (see Supp. Info. Sec.~\ref{Dicke Derivation}).

\noindent Combining Eq.~(\ref{G2 ind Gen}) and Eq.~(\ref{eq:Dicke_result}) defines the range of possible values:
\begin{equation}
\mathcal{G}^{(2)}_{\text{Ind}}(0,0) \leq
\mathcal{G}^{(2)}(0,0) \leq
\mathcal{G}^{(2)}_{\text{Dicke}}(0,0),
\label{eq:inequality}
\end{equation}
providing a natural benchmark for metasurface-mediated cooperative emission. These limits provide a benchmark against which the superradiant performance of BIC-mediated platforms can be assessed.

\emph{BIC supported superradiance}: Before exploring a specific example platform, we first present general analytical expressions that illustrate the conditions under which a BIC can produce superradiant emission, and then demonstrate these explicitly using a dielectric metasurface supporting a BIC.

For the fully inverted array, Eq.~(\ref{G2 0}) is valid to all dielectric environments, but conceals how the environment itself enables superradiance, since all environmental effects are encoded in the decay rates of Eq.~(\ref{rate}). However, in the case of a BIC-supporting metasurface, analytical insight can be gained under the following approximations. In the optimal configuration considered here, we assume that (i) each QE has the same LDOS (i.e. same single emitter decay rate $\gamma_{\mu\mu}$) and (ii) the coupling between QEs exclusively occur through the BIC mode (single-mode approximation of the metasurface) with the same coupling efficiency given by the BIC $\beta$-factor \cite{riley2025metasurfacemediatedquantumentanglementbound}; enabling the cross rate to be expressed as $\gamma_{\mu\nu} = \beta \gamma_{\mu\mu}$ for $\mu\neq\nu$. Assumption (i) is exactly true for a perfectly infinite metasurface, while assumption (ii) is always an approximation, since additional modes in the metasurface may also contribute to cross-coupling between QEs. However, these are typically non-resonant and do not support sustained coherent interactions, so for large enough emitter separations, this approximation generally improves \cite{riley2025metasurfacemediatedquantumentanglementbound}.  

By making these approximations, one can derive the second-order correlation function for $N$ emitters in the optimal configuration (see the explicit derivation of this equation in Supp. Info. Section \ref{BIC Mode G2 Derivation}):
\begin{equation}
    \mathcal{G}_{\text{BIC}}^{(2)}(0,0) = (1+\beta^2)\frac{(N-1)}{N}=\frac{(1+\beta^2)}{2}\mathcal{G}^{(2)}_{\text{Dicke}}(0,0). 
    \label{BIC only G2}
\end{equation}
This result shows that as $\beta\rightarrow 1$, $\mathcal{G}_{\text{BIC}}^{(2)}(0,0)\rightarrow \mathcal{G}^{(2)}_{\text{Dicke}}(0,0)$, i.e. this platform enables one to achieve the upper limit of Dicke superradiance, confirming that the BIC mode can sustain the desired long-range interactions required for superradiance. This result is also consistent with the result obtained in Ref.~\cite{mok2023dicke} when considering an ``all-to-all" uniform coupling of the QEs, which is in effect what the metasurface does.

Additionally, using the Lindblad master equation, a Dicke-like emission rate can be derived under the mean field and large ensemble approximations
\begin{equation}
    R_\text{BIC}(t) \approx \frac{\gamma \beta N^2}{4} \text{Sech}^2\left[ \frac{\gamma\beta N}{2} (t-t_0) \right],
\end{equation}
where the maximum value of the emission rate occurs at $t_0 = \frac{\ln[\beta N]}{\gamma\beta N}$ (Supp. Info. Section \ref{Appendix: BIC Dicke Superradiance}). When the emitters perfectly couple to the BIC mode, $\beta = 1$, one retrieves the original emission rate expression of Dicke \cite{Dicke_Paper}, again affirming the ability of the BIC to achieve maximal Dicke superradiance.

These results together show that when considering an ensemble of emitters significantly coupled to the BIC mode, one approaches the Dicke limit of Eq.~(\ref{eq:Dicke_result}), hence the BIC mode enables the generation of a quasi-Dicke superradiance within an extended 2D platform.

\emph{Metasurface-enhanced superradiance}:~We demonstrate metasurface-enhanced superradiance using the all-dielectric structure of Ref.~\cite{Diego_Paper}: a 2D square lattice of silicon scatterers (\( n = 3.5 \)) with radius \( R = 100\,\text{nm} \) and lattice constant \( a = 400\,\text{nm} \) in free space. This metasurface supports a magnetic dipole BIC at \( \lambda_\text{BIC} = 708.9\,\text{nm} \). For computational efficiency, we use a finite \( 21 \times 21 \) lattice, which sustains a high-\( Q \) quasi-BIC at the same wavelength \cite{riley2025metasurfacemediatedquantumentanglementbound}. A diagram of the metasurface and QE setup is seen in Fig.~\ref{Lattice Superraidance Diagram}.

A square array of QEs is positioned in the near field, at height \( 4\,\text{nm} \) above the surface (typical of hBN single-photon emitters \cite{do2024room}). The QE transition wavelength is resonant with the MD-BIC. Optimal coupling occurs when each QE is laterally offset by \( x_0 = 0.163\,a \) along \( x \), with dipoles oriented along \( y \), the direction of the BIC electric field \cite{riley2025metasurfacemediatedquantumentanglementbound}. This defines the \textit{optimal configuration}, with emitter positions
\(\bm{R}_\mu =[x_0+an, am, 104\,\text{nm}]\) (\( n,m \in \mathbb{Z}\)) and \( \bm{d}_\mu \parallel \hat{y} \) where $\mu \coloneqq (n,m)$. The emitter lattice constant \( d \) can differ from \( a \).


To compare free-space and metasurface-enhanced superradiance, we compute \(\mathcal{G}^{(2)}(0,0)\) [Eq.~(\ref{G2 0})] for fully inverted arrays with \(N = 3\times 3, 5\times 5, \dots, 11\times 11\) and \( d = a = 400\,\mathrm{nm}\), such that are optimally positioned. The decay rates \(\gamma_{\mu\nu}\) of the QEs are obtained using \texttt{SMUTHI} Python package \cite{SMUTHI_Paper, riley2025metasurfacemediatedquantumentanglementbound} (see Supp. Info. \ref{Appendix Methods}).

Fig.~\ref{Lattice Superraidance}a shows that the metasurface dramatically enhances \(\mathcal{G}^{(2)}(0,0)\) (red dots) compared to free space (blue dots), in good agreement with the analytical single-mode approximation (green dashed line) using the $\beta$-factor \(\beta = 81.79\%\) reported by Riley \textit{et al.} \cite{riley2025metasurfacemediatedquantumentanglementbound}. The discrepancies for larger emitter arrays are due to finite-size effects: the outer layers of emitters lie close to the edge of the finite metasurface, experiencing smaller LDOS and weaker coupling to the quasi-BIC mode, thus making approximations (i) and (ii) less valid. This finite-size effect is further amplified when doubling the emitter lattice constant to $ d=2a=800$nm (see the corresponding simulations in Supp. Info. Section~\ref{Appendix Edge Effects}). On the other hand, one can expect the agreement to improve as the metasurface size increases.

The Dicke and independent-emitter limits bound the results [Eqs.~((\ref{G2 ind Gen},\ref{eq:Dicke_result})], in alliance with inequality~(\ref{eq:inequality}). Increasing \( N \) slightly raises \(\mathcal{G}^{(2)}(0,0)\) in free space but eventually saturates, while in the metasurface case, finite-size effects cause a mild decrease for large arrays. For small lattices (e.g., \( 3 \times 3 \)), emission rate dynamics further confirm the superradiance enhancement (see Supp. Info. Section \ref{Appendix: Emission Rate}).


We next explore how metasurface-mediated long-range interactions persist for large separations by varying \( d \) for a \( 3\times 3 \) array. The central QE remains fixed at \( [x_0,0,104\,\mathrm{nm}] \) while the others are displaced radially (Fig.~\ref{Lattice Superraidance}b). In free space, cooperative emission disappears beyond \( d_\mathrm{critical} \approx 0.3\lambda_0 \) \cite{Garcia_Universal}.

In Fig.~\ref{Lattice Superraidance}b, we show the $\mathcal{G}^{(2)}(0,0)$ values computed from Eq.~(\ref{G2 0}) when the lattice constant $d$ is spanned from $d=0$ to $d=3\lambda_0$, in the presence of the metasurface (red curve) and in free space (blue curve). 
While at $d=0$, we recover Dicke's result $\mathcal{G}^{(2)}_{\text{Dicke}}(0,0)\simeq 1.78$ given by Eq.~(\ref{eq:Dicke_result}) (all the emitters are maximally coupled), the cooperative emission is drastically enhanced by the metasurface when $d = na$ with $n$ is a positive integer, which corresponds to the positions where the QEs are in the optimal configuration. 

One can see that superradiant behaviour with the metasurface can occur in arrays where emitter separations largely exceed $d_\text{critical}$. Or, viewed from a different standpoint, to achieve the same value of $\mathcal{G}^{(2)}(0,0)$, for instance $\mathcal{G}^{(2)}(0,0)\sim 1.4$ for $d=4a\simeq 2.26\lambda_0$, this would require in free space a separation $d\simeq 0.13\lambda_0$. 

These results highlight the central message of this work: the metasurface enables quasi-Dicke superradiance across \textit{macroscopically extended arrays}, far exceeding the critical separation of free-space cooperative emission, and with robustness to realistic array sizes.

\emph{Experimental uncertainty}: Up to this point, we have considered idealised arrays of emitters. To account for some of the experimental uncertainties encountered in practice, such as the difficulty in creating a perfect lattice of emitters, or to deterministically control their positions or orientations, we model these different situations to analyse the tolerance of the superradiance effect against them.

\begin{figure}[t]
\begin{minipage}{\columnwidth}
\begin{center}
\hspace*{-1mm}\includegraphics[width=0.99\textwidth]{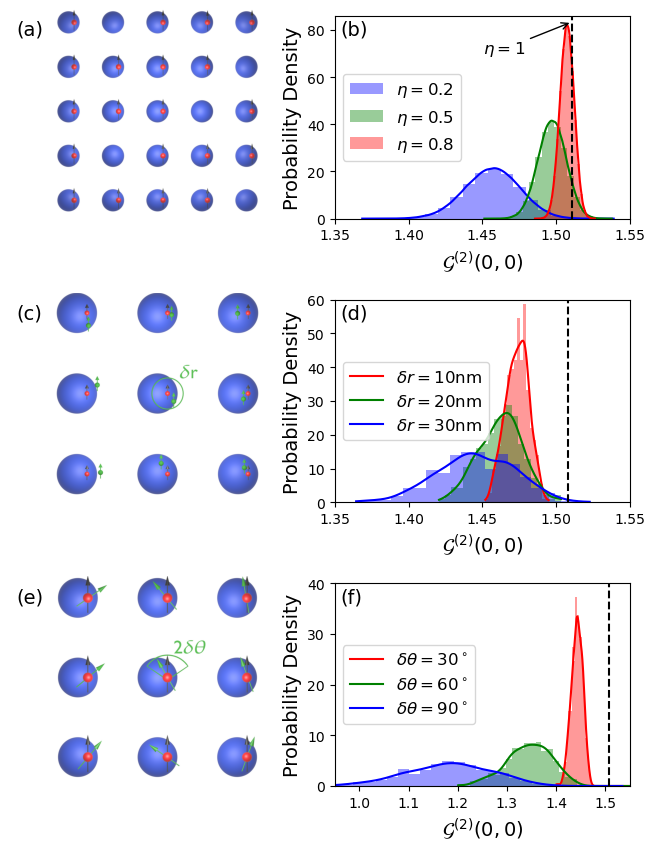}
\caption{{(a)} Incomplete lattice (\(N = 5 \times 5\)) with five missing emitters (\(\eta = 0.8\)).
{(b)} \(\mathcal{G}^{(2)}(0,0)\) distributions for \(\eta = 0.2, 0.5, 0.8\) for \(11 \times 11\) QE lattice; dashed line: full lattice (\(\eta = 1\)). 
{(c)} Positional disorder: emitters (green) displaced within radius \(\delta r\) from optimal positions (red). 
{(d)} \(\mathcal{G}^{(2)}(0,0)\) for \(\delta r = 10, 20, 30\,\mathrm{nm}\) (\(9 \times 9\) array).  {(e)} Orientation disorder: dipole angles shifted within \(\delta\theta\). {(f)} \(\mathcal{G}^{(2)}(0,0)\) for \(\delta\theta = 30^\circ, 60^\circ, 90^\circ\) (\(3 \times 3\) array); dashed line: aligned dipoles (\(\hat{\bm{y}}\)). } \label{PRL Uncert}
\end{center}
\end{minipage}
\end{figure}

{\em Incomplete QE lattices.} We first consider incomplete QEs lattices. To model them, we consider an array of $N=11\times11$ QEs with lattice constant $d=a$ in the optimal configuration, where random QEs are missing, resulting in empty spots within the lattice (see schematics in Fig.~\ref{PRL Uncert}a). We define the ``\textit{filling fraction}", denoted as $\eta$, as the fraction of QEs that have been successfully placed within the lattice, such that $\eta=0$ refers to no QEs, and $\eta=1$ means all QEs have been placed ($0\leq\eta\leq1$). 
For a given $\eta$, there are multiple possible QE configurations, each potentially resulting in a different value of $\mathcal{G}^{(2)}(0,0)$. 
A histogram of the $\mathcal{G}^{(2)}(0,0)$ distributions is shown in Fig.~\ref{PRL Uncert}b, for the values $\eta = 0.2, 0.5, 0.8$, after generating random configurations for each $\eta$ (see Supp. Info. Section \ref{Appendix: Method Experimental} for more details).

As the filling fraction decreases, the distribution of $\mathcal{G}^{(2)}(0,0)$ becomes broader, reflecting the increased number of possible emitter configurations, while the mean value decreases due to a higher occurrence of arrangements with larger mean separations between emitters and therefore weaker mutual coupling. Despite this reduction, all distributions remain well within the superradiant regime ($\mathcal{G}^{(2)}(0,0) > 1.35$), demonstrating a strong degree of robustness in the cooperative behaviour against this source of experimental noise. 

Notably, some configurations yield values exceeding the ideal case ($\eta = 1$, $\mathcal{G}^{(2)}(0,0) = 1.511$ in Fig.~\ref{Lattice Superraidance}a), which occurs when predominantly edge emitters, which experience the weakest coupling due to finite-size effects, are absent, thereby enhancing the effective collectivity of the remaining array. 

{\em Dipole position uncertainty.} Next we consider uncertainty in the position of the QEs. We restrict the positional variations within the plane parallel to the metasurface (the $xy$-plane), i.e. to 2D instead of 3D, since the uncertainty in the out-of-plane direction can be minimised, for example, if the QEs are confined within a few-layer-thick 2D material, such as in hexagonal boron nitride \cite{do2024room}.

To model this, we consider an array of $N=3\times 3$ QEs with lattice constant $d = a$, and we introduce a random shift in the \(x\)- and \(y\)-directions for each QE from the optimal position, taking any values from $0$ to a maximum shift \(\delta r\) (as shown schematically in Fig.~\ref{PRL Uncert}c). The dipole moment orientation of each QE remains fixed (aligned with the $y$-axis).

Similarly, for a given maximum shift \(\delta r\), there are multiple different configurations that can potentially result in different \(\mathcal{G}^{(2)}(0,0)\) values. A histogram showing the respective \(\mathcal{G}^{(2)}(0,0)\) distributions is shown in Fig.~\ref{PRL Uncert}d, for the various maximum shift values of \(\delta r = [10, 20, 30] \, \text{nm}\), after generating random configurations for each $\delta r$ (see Supp. Info. Section \ref{Appendix: Method Experimental}). These shifts are considered to be the typical uncertainties that one could reach experimentally, using, for example, colloidal quantum dots \cite{D4NR02123A}.

Again, the variations of the peak \(\mathcal{G}^{(2)}(0,0)\) values and the spread are remarkably small (the horizontal axis range is the same as in Fig.~\ref{PRL Uncert}b to facilitate comparison), showing that superradiant behaviour is still preserved.

{\em Dipole orientation uncertainty.} Finally, we analyse uncertainty in the dipole moment orientation. We restrict again the orientation uncertainty to the plane parallel to the metasurface (the $xy$-plane), i.e., 2D instead of 3D, which can be the case, for example, in engineered perovskite nanocrystals or nanoplatelets \cite{jurow2019manipulating, marcato2022confinement}. 

To model this situation, we similarly consider an array of $N=3\times 3$ QEs with lattice constant $d = a$, but now the QE positions are fixed and the dipole orientation is allowed to randomly vary from $0$ to a maximum angle shift $\delta\theta$, which characterizes the maximum magnitude of the in-plane angle shift possible from the optimal configuration (as shown schematically in Fig.~\ref{PRL Uncert}e). For a given maximum angle shift $\delta\theta$, multiple different configurations are possible resulting potentially in different \(\mathcal{G}^{(2)}(0,0)\) values. Since dipole orientation is arguably the most difficult parameter to control experimentally, we consider three different maximum shifts $\delta\theta = [30^\circ, 60^\circ, 90^\circ]$ up to total uncertainty of the in-plane dipole moment orientation (\(\delta\theta = 90^\circ\)).
A histogram showing the resulting distributions is shown in Fig.~\ref{PRL Uncert}f.

The mean values of \(\mathcal{G}^{(2)}(0,0)\) decrease more drastically with the maximum angle shift than for the other type of uncertainties considered thus far (note that the horizontal axis in Fig.~\ref{PRL Uncert}f is different from Fig.~\ref{PRL Uncert}b and d). However, and rather remarkably, superradiance is still present, as indicated by \(\mathcal{G}^{(2)}(0,0)\) values consistently greater than $1$, even for a total uncertainty (\(\delta\theta = 90^\circ\)). On the other hand, if a certain level of control of the orientation is present, \(\mathcal{G}^{(2)}(0,0)\) can approach the ideal case (in vertical dashed black line). Such a level of control could be in reach in the near future \cite{marcato2022confinement}.

\emph{Summary}.
We have shown that delocalized bound states in the continuum provide a distinct mechanism for mediating Dicke-like superradiance across extended two-dimensional emitter arrays, even when inter-emitter separations greatly exceed the emission wavelength. In this regime, cooperative enhancement is governed by the coupling efficiency to the BIC, quantified by the metasurface \(\beta\)-factor, and approaches the ideal Dicke limit as \(\beta \to 1\). Unlike free-space or cavity-mediated schemes, the collective response is sustained by the nonlocal character of the BIC mode rather than subwavelength spacing or mode confinement. Importantly, the cooperative emission remains robust against incomplete lattices, positional disorder, and dipole misalignment, underscoring the intrinsic resilience of BIC-mediated interactions. These results establish BIC metasurfaces as a scalable and controllable platform for realizing collective light--matter interactions beyond conventional spatial constraints, with direct implications for compact superradiant sources and cooperative quantum-photonic devices.

\bibliographystyle{apsrev4-2}
\bibliography{biblio}
\newpage

\appendix
\begin{widetext}

\section*{\Large Supplementary Information: Dicke Superradiance in Extended 2D Quantum Arrays Coupled to Metasurface
Bound States in the Continuum}

\section{Collective mode equation in a generalised environment} \label{Garcia Collective Equation}
As discussed within the main text, when considering the fully inverted array, the limitations of the dimensionality of the Hilbert space can be overcome through Eq.~(\ref{G2 0}). However, another issue arises due to the nested iteration, meaning the calculation has a time complexity of $\mathcal{O}(N^4)$ and so is not practical when considering large ensembles of QEs. Work conducted by Asenjo-Garcia et al. \cite{Garcia_Universal} provides an optimised approach to compute this through the eigenvalues $\bm{\Gamma}|_{\mu\nu} = \gamma_{\mu\nu}$. However, the work conducted is limited to the case in which all QEs experience the same LDOS ($\gamma_{\mu\mu} = \gamma$ $\forall \mu$), and so in this section, the derivations, based on that conducted by Asenjo-Garcia et al., will be shown to obtain Eq.~(\ref{G2 Garcia Generalised}). 

To achieve this generalisation, we begin by decomposing $\bm{\Gamma}$ into its eigenvalues and eigenvectors, $\gamma_{\mu\nu} = \sum_i \Gamma_i \alpha_{i,\mu} \alpha_{i,\nu}^*$, where $\alpha_{i,\mu}$ is the $\mu^\text{th}$ component of the $i^\text{th}$ eigenvector. This decomposition can be substituted into Eq.~(\ref{G2 0}) producing 
\begin{equation}
\begin{split}
    \mathcal{G}^{(2)}(0,0) = \frac{\sum_{ij} \Gamma_i \Gamma_j \sum_{\mu\nu\gamma\epsilon} \alpha_{i,\epsilon} \alpha_{i,\mu}^* \alpha_{j,\gamma} \alpha_{j, \nu}^* (1-\delta_{\mu\nu})(\delta_{\mu\epsilon}\delta_{\gamma\nu} + \delta_{\mu\gamma}\delta_{\nu\epsilon}) }{(\sum_i \Gamma_i)^2}.
\end{split}
\end{equation}
Note that we have used the relation $\sum_\mu \gamma_{\mu\mu} = \text{Tr}[\bm{\Gamma}] = \sum_i \Gamma_i$. From here, the brackets can be expanded to produce
\begin{equation}
\begin{split}
    \mathcal{G}&^{(2)}(0,0) = \frac{1}{(\sum_i \Gamma_i)^2}\sum_{ij} \Gamma_i \Gamma_j \left[ \delta_{ij} + 1 - 2\sum_\mu |\alpha_{i,\mu}|^2 |\alpha_{j,\mu}|^2 \right]
\end{split}
\end{equation}
where the orthonormal nature of the eigenvectors has been employed, $\sum_\mu \alpha_{i,\mu} \alpha_{j,\mu}^* = \bm{\alpha}_i\cdot\bm{\alpha}_j = \delta_{ij}$. Upon a final expansion of Eq.~(\ref{G2 Garcia Generalised}) is obtained
\begin{equation}
\small
    \mathcal{G}^{(2)}(0,0) = 1 + \frac{\sum_\mu^N \Gamma_i^2}{\big(\sum_i^N \Gamma_i\big)^2} - \frac{2 \sum_\mu^N \Big[ \big( \sum_i^N \Gamma_i |\alpha_{i,\mu}|^2 \big)^2 \Big]}{\big(\sum_i^N \Gamma_i\big)^2}.
    \label{G2 Garcia Generalised}
\end{equation}
Within Ref.~\cite{Garcia_Universal}, they employ the relation $\sum_{\nu} A_\nu |\alpha_{\nu, i}|^2 = \langle A_\nu \rangle$; however, this is only valid when all diagonal elements are identical within the matrix, which is not the case in general.

\section{Second-order correlation functions: limiting cases}

\subsection{Independent emitters} \label{Appendix Independent emitters}
The first system considered is a collection of $N$ independent emitters, such that the collective decay rates are identically zero. In terms of the decay matrix, $\bm{\Gamma}$, all off-diagonal elements are zero, so the decay matrix can be defined as $\Gamma_{\mu\nu} = \gamma_{\mu\nu} \delta_{\mu\nu}$, where $\delta_{\mu\nu}$ is the Kronecker delta. In the case of a homogeneous environment, all off-diagonal elements are identical, $\gamma_{\mu\mu} = \gamma \ \forall \mu\in[1,N]$, but in the case of an arbitrary environment, this is not the case in general. So, retaining this generalisation, substituting this expression into equation \ref{G2 0} results in 
\begin{equation}
    \mathcal{G}_{\text{Ind}}^{(2)}(0,0) = \frac{\sum_{\mu\nu\gamma\epsilon}^N \gamma_{\epsilon\mu}\gamma_{\gamma\nu} \delta_{\epsilon\mu}\delta_{\gamma\nu} (1-\delta_{\mu\nu})(\delta_{\mu\epsilon}\delta_{\nu\gamma} + \delta_{\mu\gamma}\delta_{\nu\epsilon})}{\Big(\sum_{\mu\nu}^N \gamma_{\mu\nu}\delta_{\mu\nu} \Big)^2}
\end{equation}
Upon expanding the bracket and reducing the indices through the use of the Kronecker delta terms, we obtain the expression 
\begin{equation}
\begin{split}
    \mathcal{G}_{\text{Ind}}^{(2)}(0,0) &= \frac{(\sum_{\mu}^N \gamma_{\mu\mu})^2 - \sum_{\mu}^N \gamma_{\mu\mu}^2}{(\sum_{\mu}^N \gamma_{\mu\mu})^2} \\
    &= 1 - \frac{\sum_{\mu}^N \gamma_{\mu\mu}^2}{(\sum_{\mu}^N \gamma_{\mu\mu})^2}.
\end{split}
\end{equation}
Which is the generalised expression for independent emitters within an arbitrary electromagnetic environment. The evident result from this is that as the sum of squared terms, $\sum A^2$, will always be less than the square of the sum, $(\sum A)^2$, this equation is bound by the range $0 < \mathcal{G}_{\text{Ind}}^{(2)}(0,0) < 1$, showing that independent emitters cannot produce superradiant behavior within any electromagnetic environment. But this result could be highly useful when trying to general strong anti-bunching behavior, as this expression can result in a lower bound than the case of a homogenous environment, as seen in Figure \ref{Lattice Superraidance}c. 

To ensure consistency from this equation, in the case of a homogenous environment, where $\gamma_{\mu\mu} = \gamma$, we achieve the result
\begin{equation}
\begin{split}
    \mathcal{G}_{\text{Ind,Hom}}^{(2)}(0,0) &= 1 - \frac{\sum_{\mu}^N \gamma^2}{(\sum_{\mu}^N \gamma)^2} \\
    & = 1 - \frac{\gamma^2\sum_{\mu}^N 1}{\gamma^2 (\sum_{\mu}^N 1)^2}  = 1- \frac{N}{N^2} = \frac{N-1}{N}
\end{split}
\end{equation}
which is the commonly seen result for independent emitters within a homogeneous environment.

\subsection{Dicke scenario} \label{Dicke Derivation}
We now consider the scenario in which all emitters maximally couple to each other. This is the scenario initially considered by Dicke, but this is limited to the case in which all emitters experience the same LDOS. Here, this result will be generalised to account for varying LDOS for each emitter. 

To do this, a Cauchy-Schwarz-like inequality for the CDOS and LDOS is employed, which states 
\begin{equation}
\begin{split}
    &\text{Im}[\bm{d}_1\cdot\bm{G}(\bm{r}_1, \bm{r}_2)\cdot\bm{d}_2] \leq \\
    & \sqrt{\bm{d}_1\cdot\text{Im}[\bm{G}(\bm{r}_1, \bm{r}_1)\cdot\bm{d}_1] ~~\text{Im}[\bm{d}_2\cdot\bm{G}(\bm{r}_2, \bm{r}_2)]\cdot\bm{d}_2},
\end{split}
\end{equation}
from which follows that the maximum possible value for the cross decay rate is given by $\gamma_{12} = \sqrt{\gamma_{11}\gamma_{22}}$. This expression can be inserted into Eq.~(\ref{G2 0}) produces
\begin{equation}
\begin{split}
    \mathcal{G}_{\text{Dicke}}^{(2)}(0,0) &= \frac{2(\sum_{\mu}^N \gamma_{\mu\mu})^2 - 2\sum_{\mu}^N \gamma_{\mu\mu}^2}{(\sum_{\mu}^N \gamma_{\mu\mu})^2} = 2- \frac{2\sum_{\mu}^N \gamma_{\mu\mu}^2}{\big(\sum_{\mu}^N \gamma_{\mu\mu}\big)^2} = 2\mathcal{G}_{\text{Ind}}^{(2)}(0,0).
\end{split}
\end{equation}
In the specific scenario of a homogeneous environment or when all emitters occupy the same position in space, as originally considered by Dicke, this result simplifies to 
\begin{equation}
    \begin{split}
        \mathcal{G}_{\text{Dicke,H}}^{(2)}(0,0) = \frac{2(N-1)}{N}.
    \end{split}
\end{equation}
This is the result originally obtained by Dicke.

\subsection{True BIC} \label{BIC Mode G2 Derivation}
The final correlation function we will derive here is when considering $N$ emitters in the presence of a metasurface, which supports a BIC at the transition wavelength of the emitters. As discussed in the main text, this calculation is, in general, not tractable analytically due to the inhomogeneous nature of the metasurface, and in turn, the LDOS is spatially varying. To avoid this complexity, we consider all emitters to reside at the same position within a metasurface unit cell, so all emitters experience the same LDOS (which we chose to correspond to the maximum LDOS value available).

Additionally, the BIC mode is not the only mode supported by the metasurface, and so the influence of these additionally leaky modes will influence the calculation. Here we consider only the coupling of the emitters through the BIC mode. This coupling can be characterized by the $\beta$ factor for the setup described previously, such that $0\leq\beta\leq$. For a given value of $\beta$, the collective decay rate between any two emitters is defined as 
\begin{equation}
    \gamma_{\mu\nu} = \begin{cases}
        \gamma \ \ \ \ \ \text{for $\mu = \nu$} \\
        \beta \gamma \ \ \ \text{for $\mu \not= \nu$}. \\
    \end{cases}
\end{equation}

where $\gamma$ is the spontaneous emission rate for a single emitter at the optimal position in the presence of the metasurface. To obtain an expression similar to those seen previously, we must expand the Kronecker delta terms before substituting the expression for the decay matrix, due to the altered definition.
\begin{equation}
\begin{split}
    \mathcal{G}^{(2)}_{\text{BIC}}(0,0) &= \frac{\sum_{\mu\nu}^N \gamma_{\mu\mu}\gamma_{\nu\nu} + \sum_{\mu\nu}^N \gamma_{\mu\nu}\gamma_{\nu\mu} - 2\sum_{\mu}^N \gamma^2_{\mu\mu} }{N^2 \gamma_0^2} \\
    &= \frac{N^2 \gamma^2 - 2N\gamma^2 + \sum_{\mu\nu}^N \gamma_{\mu\nu}\gamma_{\nu\mu}}{N^2\gamma^2}.
\end{split}
\end{equation}
The final summation term must be first split into terms for $\mu = \nu$ and $\mu \not= \nu$ to enable the use of the prior relation, which yields
\begin{equation}
\begin{split}
    \mathcal{G}^{(2)}_{\text{BIC}}(0,0) &= \frac{N^2 \gamma^2 - 2N\gamma^2 + N\gamma^2 + \sum_{\mu\not=\nu}^N \gamma_{\mu\nu}\gamma_{\nu\mu}}{N^2\gamma^2} \\
    & = \frac{N^2 \gamma^2 - 2N\gamma^2 + N\gamma^2 + \sum_{\mu\not=\nu}^N (\beta^2 \gamma^2) }{N^2\gamma^2} \\
    & = \frac{N^2 \gamma^2 - N\gamma^2  + (N^2-N)(\beta^2 \gamma^2) }{N^2\gamma^2} \\
    & = \frac{(1+\beta^2)(N-1)}{N}.
\end{split}
\end{equation}
As can be seen from this result, in the limit of either $\beta=1$ or $\beta = 0$, the Dicke and independent emitters results are obtained, respectively. We emphasize that this model does not consider the effects of the additional models present in the metasurface. However, in cases of high $\beta$ factors, this implies the influence of these other modes is less significant and in turn, implies a higher accuracy of this model's predictions.

\begin{figure}[t]
\begin{minipage}{\columnwidth}
\begin{center}
\hspace*{-1mm}\includegraphics[width=0.5\textwidth]{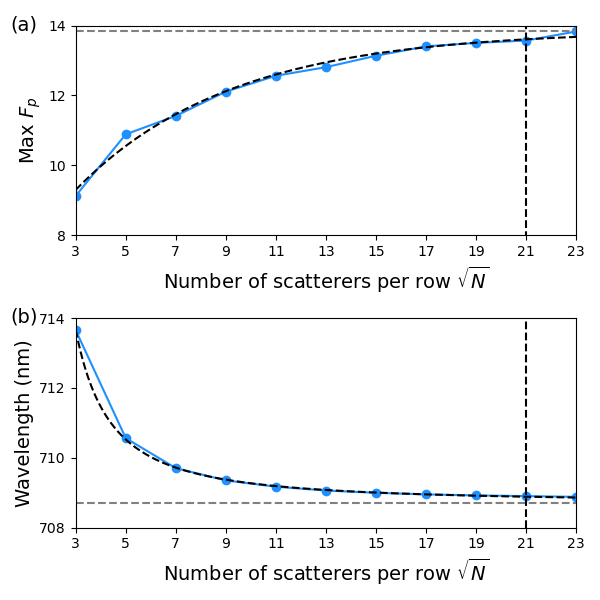}
\caption{Plots showing the convergence of the maximum value of the Purcell Factor \textbf{(a)} Max[$F_P$] and \textbf{(b)} resonant Wavelength as a function of the number of scatterers included in the square lattice. The dashed lines in both plots are numerical fittings using the expressions (a) $\text{Max}[F_p] = F_{p,\infty} + A e^{-B*N}$ and (b) $\lambda = \lambda_\infty + C/N^p$, shown by the grey dashed curves. The infinite limit values, $F_{p,\infty}=13.85$ and $\lambda_\infty=708.71$nm, are indicated by the horizontal dashed line in (a) and (b), respectively. The vertical dashed line indicates the $21\times21$ lattice considered throughout the main text. }\label{BIC Convergance Appendix Plot}
\end{center}
\end{minipage}
\end{figure}

\section{Symmetric-projection derivation of the metasurface-enhanced emission rate} \label{Appendix: BIC Dicke Superradiance}

We derive a closed form for the emission rate of an ensemble of $N$ identical two-level emitters that decay both collectively (via a BIC mode) and independently. The dynamics is taken to be Markovian and to conserve the total excitation number under the system Hamiltonian; the environment-induced decay is modelled by the Lindblad equation with a collective channel weighted by the BIC coupling efficiency $\beta\in[0,1]$ and an independent channel accounting for residual (non-collective) emission:
\begin{equation}
\dot{\rho} = -i[H_S,\rho] + \gamma\beta\,\mathcal{D}[S_-]\rho + \gamma(1-\beta)\sum_{\mu=1}^N\mathcal{D}[\sigma_-^{(\mu)}]\rho,
\label{eq:lindblad}
\end{equation}
where $\mathcal{D}[L]\rho = L\rho L^\dagger-\tfrac{1}{2}\{L^\dagger L,\rho\}$, 
$S_\pm=\sum_\mu\sigma_\pm^{(\mu)}$, and $\gamma$ denotes the single-emitter decay rate at the optimal position in the presence of the metasurface.

Introduce the projector $\Pi_n$ onto the manifold with exactly $n$ excitations and the manifold populations $P_n(t)\equiv\mathrm{Tr}[\Pi_n\rho(t)]$. Using standard projector identities and projecting \eqref{eq:lindblad} onto the $n$-manifold yields the birth--death chain
\begin{align}
\dot{P}_n &= \gamma\beta\Big[(N-n)(n+1)P_{n+1}-n(N-n+1)P_n\Big] \notag + \gamma(1-\beta)\Big[(n+1)P_{n+1}-nP_n\Big].
\label{eq:ladder}
\end{align}

Define the mean excitation number
\begin{equation}
n(t)\equiv\sum_{n=0}^N n\,P_n(t)=\langle\hat n\rangle_t.
\label{eq:mean-excitation}
\end{equation}
Multiplying \eqref{eq:mean-excitation} by $n$ and summing over $n$ gives
\begin{equation}
\dot{n} = -\gamma\beta\langle n(N-n+1)\rangle_t - \gamma(1-\beta)\,n(t).
\label{eq:ndot-exact}
\end{equation}
To close \eqref{eq:ndot-exact} we adopt the mean-field approximation valid near the superradiant burst and for large $N$,
\begin{equation}
\langle n(N-n+1)\rangle_t \approx n(t)[N-n(t)+1].
\label{eq:mf-closure}
\end{equation}
This reduces \eqref{eq:ndot-exact} to the nonlinear ODE
\begin{equation}
\dot{n} = -\gamma(1+\beta N)\,n + \gamma\beta\,n^2.
\label{eq:ndot-mf}
\end{equation}

The emission rate is $R(t)\equiv -\dot{n}(t)$. Solving \eqref{eq:ndot-mf} with $n(0)=N$ gives
\begin{equation}
R(t)=\frac{\gamma(1+\beta N)^2}{4\beta}\,\mathrm{sech}^2\!\left[\tfrac{\gamma(1+\beta N)}{2}(t-t_0')\right],
\label{eq:emission-general}
\end{equation}
with
\begin{equation}
t'_0=\frac{\ln(\beta N)}{\gamma(1+\beta)N}.
\label{eq:t0-general}
\end{equation}
Equation~\eqref{eq:emission-general} reproduces the expected limits: $\beta\to 0$ yields the exponential independent decay $R(t)=\gamma N e^{-\gamma t}$, while $\beta\to 1$ and large $N$ recover Dicke’s quadratic scaling.

For regimes with $\beta N\gg1$ one may further approximate $1+\beta N\approx\beta N$ in Eqs.~\eqref{eq:emission-general} and \eqref{eq:t0-general}, yielding the simplified quasi-Dicke form used in the main text,
\begin{equation}
R_{\mathrm{BIC}}(t)\approx\frac{\gamma\beta N^2}{4}\,\mathrm{sech}^2\!\left[\tfrac{\gamma\beta N}{2}(t-t_0)\right],
\label{eq:emission-final}
\end{equation}
with
\begin{equation}
t_0=\frac{\ln(\beta N)}{\gamma\beta N},
\end{equation}
which corresponds to Eq.~(10) of the main text.

\section{Simulation methods}
\label{Appendix Methods}

\subsection{General}
In this appendix, the procedures used to generate the results seen in the main text are explored. To do this, the method will be separated into two sections, corresponding to Fig.~\ref{Lattice Superraidance} and Fig.~\ref{PRL Uncert}. All of the following calculations employ the Python package SMUTHI to compute the classic electromagnetic results \cite{SMUTHI_Paper} . Due to the employment of dipole sources, the periodicity can not be employed, and so, a finite $21\times21$ lattice of spherical scatterers is generated in the $x$-$y$ plane. This size results from convergence testing, seen in Fig.~\ref{BIC Convergance Appendix Plot}, ensuring sufficiently converged results, whilst maintaining tractable calculations. Within SMUTHI, one must set the multipole order of the scatterers, and in this work, it is limited to one, as it results in good agreement with calculations done using COMSOL \cite{comsol64}. 

To compute the spontaneous emission rate, $\gamma_{\mu\mu}$ for an emitter positioned at $\bm{R}$ in the presence of the metasurface, the power dissipated by the dipole is calculated, $P$, and is then normalised by the power dissipated in free space, $P_0$, at an identical frequency, yielding the Purcell factor, $F_p$. This is then multiplied by the analytical expression for the free space decay rate, $\gamma_0$. This approach is used to obtain a value within the desired unit, as SMUTHI adheres to a "relative units philosophy", meaning the values outputted are not in standard SI form. This can be summarised in the expression 
\begin{equation}
    \gamma_{\mu\mu}(\bm{R},\omega_0) = \frac{P(\bm{R},\omega_0)}{P_0(\omega_0)} \gamma_0 (\omega_0)
\end{equation}

To compute the cross decay rate of an emitter, $\gamma_{\mu\nu}$, between a emitters positioned at $\bm{r}_1$ and $\bm{r}_2$ and posessing dipole moments $\bm{d}_1$ and $\bm{d}_2$ respectively, a dipole source is placed at $\bm{r}_1$ and the electric field is evaluated at $\bm{r}_2$, this resulting electric field is then inserted into the expression 
\begin{equation}
    \gamma_{\mu\nu} = \frac{2}{\epsilon_0\hbar c^2} \text{Im}[\bm{d}_2 \cdot \bm{E}(\bm{r}_2)].
\end{equation}
Using this approach, the desired dipole moment $\bm{d}_1$ must be multiplied by the factor $1/8.85\times10^{-39}$ and the positions must be input in nanometers to yield electric field values in SI units. This approach can be inverted by placing a source dipole at $\bm{r}_2$ and evaluating the electric field at $\bm{r}_2$, yielding an identical result.

\subsection{Metasurface-enhanced superradiance}

Having established how the desired decay rates are computed using SMUTHI, Fig.~\ref{Lattice Superraidance}a and Fig.~\ref{Lattice Superraidance}b can be generated. 

\textit{Number of emitters}: To obtain Fig.~\ref{Lattice Superraidance}a, we require the full decay matrix for each system of $N\times N$ emitter lattice to enable the calculation of $\mathcal{G}^{(2)}(0,0)$. Each emitter is located at $\bm{R}_\mu = [x_0 + a(i-(N-1)/2), a(j-(N-1)/2), 104\text{nm}]$, with $0 \leq i,j < N$ are iteration parameters to generate the full set of positions, $\mu \equiv (i,j)$ and all dipole moments are parallel to the $\hat{\bm{y}}$. Following this, the procedure outlined previously is employed to compute the different decay rates. To optimise this calculation, the symmetry $\gamma_{\mu\nu} = \gamma_{\nu\mu}$ can be used to reduce the number of calculations required. 

Additionally, a further method to optimise the calculation can be to note that the decay matrix of an $n\times n$ emitter lattice is simply a subsample of a lattice of $m\times m$, where $n<m$. So, only the full decay matrix of $11\times11$ is required to be computed, and then to generate $\mathcal{G}^{(2)}(0,0)$ for smaller lattices, a systematic subsampling is used to generate the corresponding decay matrix. Once the decay matrix is obtained, Eq.~(\ref{G2 0}) is used to compute $\mathcal{G}^{(2)}(0,0)$. 

\textit{Emitter separation}: A similar approach can be employed to compute the results seen in Fig.~\ref{Lattice Superraidance}b. The emitter configuration in this case is $\bm{R}_\mu = (x_0 + di, dj, 104\text{nm})$, where $d$ is the emitter lattice constant and the indices are bound by $i,j \in [-1,0,1]$. The parameter $d$ is discretized from $0$ to $3\lambda$, and the decay matrix for each value of $d$ is computed, and in turn $\mathcal{G}^{(2)}(0,0)$ is computed.

\section{Finite-size effects} \label{Appendix Edge Effects}

As discussed within the main text, when considering a finite metasurface, edge effects can play a significant role when considering emitters positioned away from the centre of the metasurface. This effect can already be seen in Fig.~\ref{Lattice Superraidance}b, where, as the emitter separations increase and approach the edge of the metasurface, the value of $\mathcal{G}^{(2)}(0,0)$ decreases. We can show this effect even more clearly through an additional plot introduced into Figure \ref{Lattice Superraidance}b, where we consider the emitter separation to now be 800nm, as opposed to the currently used value of 400nm. This increased separation leads to the edge emitters in the case of $11\times11$ emitter lattice being directly above the edge spherical scatterers (with the additional offset to maximally couple to the MD-BIC mode), and so edge effects are expected to be maximal. Fig.~\ref{800nm separation plot} shows the results from Fig.~\ref{Lattice Superraidance}a, now with the introduction of an emitter lattice with an emitter separation of $d=800$nm. As is expected, for small systems, there are consistent results between the $800$nm and $400$nm results, as edge effects are minimal. However, when considering larger emitter lattices, the values of $\mathcal{G}^{(2)}(0,0)$ begin to diverge, and this is resulting from the edge effects becoming much more prominent in the $800$nm case, due to the emitter being closer to the fringe of the metasurface.

\section{Metasurface-enhanced emission rate} \label{Appendix: Emission Rate}

\begin{figure}[t]
\begin{minipage}{\columnwidth}
\begin{center}
\hspace*{-1mm}\includegraphics[width=0.5\textwidth]{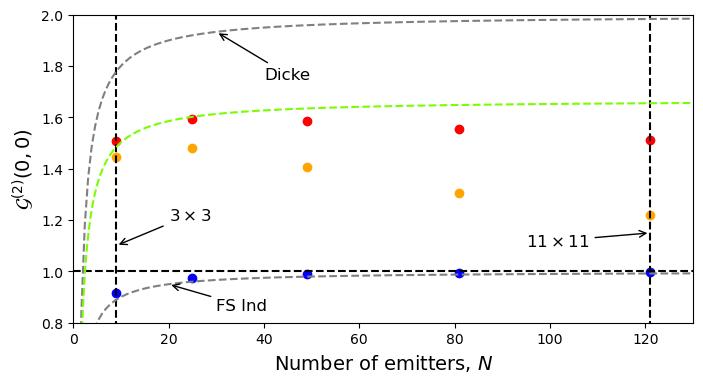}
\caption{ $\mathcal{G}^{(2)}(0,0)$ as a function of the total number of emitters for the finite metasurface (red), infinite metasurface under the single mode approximation in Eq.~(\ref{BIC only G2})(green), and in free space (blue). As seen previously in Fig.~\ref{Lattice Superraidance}a, but now with the addition of an emitter lattice with separation $d=800$nm in the presence of the metasurface.}\label{800nm separation plot}
\end{center}
\end{minipage}
\end{figure}

A key signature of superradiant emission is a peak in the emission rate occurring at some time $t_{max} > 0$. The emission rate of a given ensemble of QE is defined as
\begin{equation}
\begin{split}
    R(t) &= \sum_{\mu\nu}^N\gamma_{\mu\nu} \braket{\sigma_\mu^+(t)\sigma_\nu^-(t)} \\
    & = \sum_{\mu\nu}^N \gamma_{\mu\nu} \text{Tr}[\sigma_\mu^+\sigma_\nu^- \rho(t)],
\end{split}
\end{equation}
where $\rho(t)$ is the density matrix of the QE array, and the cyclical property of the trace operation has been used. To obtain a solution for $R(t)$, a master equation is required to determine the evolution of $\rho(t)$. Here, the system is assumed to be weakly coupled to the environment, allowing the use of the Lindblad master equation
\begin{figure}[t]
\begin{minipage}{\columnwidth}
\begin{center}
\hspace*{-1mm}\includegraphics[width=0.5\textwidth]{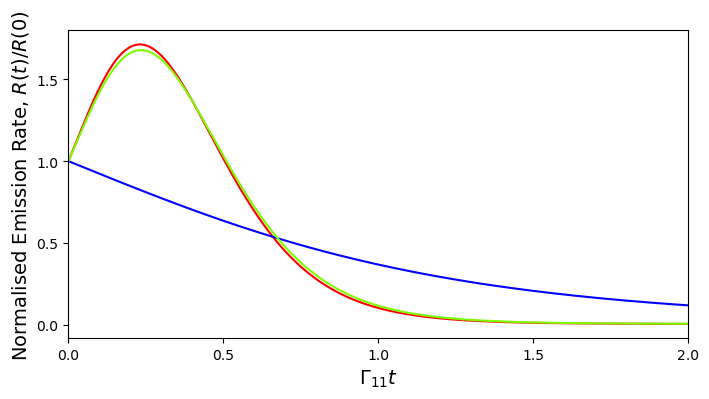}
\caption{The normalised emission rate of 9 QE arranged into a square lattice with an emitter lattice constant of $a=400$nm with dipole moments parallel to the y-axis. The electromagnetic environments considered are the finite silicon scatterer metasurface (red), infinite silicon scatterer metasurface with $\beta = 81.79$\% (green), and free space (blue) as considered in Fig.~\ref{Lattice Superraidance}b. The dynamics of the QE lattice is governed by the Lindblad master equation in \ref{Master equation}, with an initial state being the fully inverted array, $\rho(0) = \ket{e}\bra{e}^{\otimes N}$.}\label{Emission Rate Plot}
\end{center}
\end{minipage}
\end{figure}
\begin{equation}
    \begin{split}
        \Dot{\rho}(t) =& -i\sum_\mu^N \omega_0 [\sigma_\mu^+ \sigma_\mu^-, \rho(t)] + \sum_{\mu\not=\mu}^N \Delta_{\mu\nu} [\sigma_\mu^+ \sigma_\nu^-, \rho(t)] + \sum_{\mu\nu}^N \gamma_{\mu\nu} \Big( \sigma_\mu^-\rho(t)\sigma_\nu^+ - \frac{1}{2} \{\sigma_\nu^+\sigma_\mu^-,\rho(t)\} \Big).
    \end{split}
    \label{Master equation}
\end{equation}
where $\Delta_{\mu\nu} = (\mu_0 \omega_0^2/\hbar)~\text{Re}[\bm{d}^*_\mu\cdot\bm{G}(\bm{R}_\mu, \bm{R}_\nu, \omega_0)\cdot \bm{d}_\nu]$ is the dipole coupling rate and ($\{A,B\} = AB + BA$) $[A,B] = AB-BA$  is the (anti-) commutator. This expression assumes the environment is at a temperature $T=0K$, such that no thermal excitations are present. Solving this equation via the Python package QuTiP \cite{Qutip} , and inserting the resulting density matrices at each time step, yields the emission rate, as seen in Fig.~\ref{Emission Rate Plot} for both free space and when coupled to the BIC mode.

\section{Experimental uncertainty} \label{Appendix: Method Experimental}
The results seen in Fig.~\ref{PRL Uncert} employ a similar procedure to that discussed previously within Section \ref{Appendix Methods}, with only minor alterations.

To obtain the results seen in Fig.~\ref{PRL Uncert}a, the decay matrix for the total system of $11\times11$ emitters is computed, as done when computing Fig.~\ref{Lattice Superraidance}a, and then for a given filling fraction value, $\eta$, a random sample of emitters are taken, and the corresponding decay matrix is generated and $\mathcal{G}^{(2)}(0,0)$ is calculated for the sample. This process is repeated for $\mathcal{O}(10^4)$ different random samples until a sufficient number is taken to achieve a converged distribution of $\mathcal{G}^{(2)}(0,0)$. This process is again repeated for the different values of $\eta$. 
The mean value $\mu_\eta$, standard deviation $\sigma_\eta$, and skewness, $\gamma_\eta$ are computed using the expressions 
\begin{subequations}
    \begin{gather}
        \mu_\eta = \frac{1}{n} \sum_i^n \mathcal{G}^{(2)}_{\eta,i}(0,0), \label{Mean}\\
        \sigma_\eta = \sqrt{\frac{1}{n-1} \sum_i^n \Big(\mathcal{G}^{(2)}_{\eta,i}(0,0) - \mu_\eta\Big)^2 } \label{Std} \\
        \gamma_\eta = \frac{1}{(n-1)\sigma_\eta^3}\sum_{i}^{n} \Big(\mathcal{G}^{(2)}_{\eta,i}(0,0) - \mu_\eta\Big)^3 
    \end{gather}
    \label{Stat eqns}
\end{subequations}
where $n$ is the number of random samples taken and $\mathcal{G}^{(2)}_{\eta,i}(0,0)$ is the value of the second-order correlation function of the $i^\text{th}$ sample for a given $\eta$ value. As we are generating a set of random samples, the sample variance was employed. 

To compute the positional uncertainty seen in Fig.~\ref{PRL Uncert}b, the $3\times3$ lattice of emitters begins at their optimal positions, $\bm{R}_\mu = (x_0 + ai, aj, 104\text{nm})$. For a given value of maximal offset $\delta r$, a set of random offsets is generated, $[\delta x_1, \delta y_1, \delta x_2, \delta y_2, ... \delta x_9, \delta y_9]$ which correspond to the offset of the dipole positioning in the x-y plane and obey the condition $\sqrt{\delta x_\mu^2 + \delta y_\mu^2} \leq \delta r$. This results in each emitter having the new positioning $\bm{R}'_\mu = (x_0 + ai + \delta x_\mu , aj + \delta y_\mu, 104\text{nm})$. For this new set of positions $\mathcal{G}^{(2)}(0,0)$ is computed, and the process is repeated for $\mathcal{O}(10^3)$ random displacement values until a converged result is obtained. This is then repeated for different values of $\delta r$. This computation is complex, and so a simplification was employed to reduce the number of calculations; the random displacements are discretised to 100 possible values, allowing one to track previously calculated decay rates, and so if a calculation is repeated in a later sample, the calculation can be skipped, resulting in a more efficient run. This discretisation results in a step size of $2\delta r/ 50$, which is 0.4, 0.8 and 1.5 for $\delta r = [10, 20, 30]$nm, respectively. The dipole moment of each dipole in this procedure is fixed parallel to the y-axis

For the dipole uncertainty, a near identical process is used, but in this case, a single random parameter is used, $\theta_\mu$, which characterises the in-plane orientation shift in the dipole moment. This parameter is bound by $|\theta_\mu| \leq \delta \theta$, and a set of 9 random values are generated, and applied to the dipole moments, such that the dipole moments become $\bm{d}_\mu = d [\cos(90^\circ + \theta_\mu), \sin(90^\circ + \theta_\mu), 0]$. Using these shifted moments, the decay matrix can be computed, and in turn $\mathcal{G}^{(2)}(0,0)$. This process is repeated $\mathcal{O}(10^3)$ times, until a converged disribution of $\mathcal{G}^{(2)}(0,0)$ is obtained. Again, the parameter $\theta_\mu$ is discretised, allowing the tracking of previous decay rate calculations to prevent repeated runs. Here, the discretisation was done over 100 steps, yielding step sizes of $[0.6^\circ, 1.2^\circ, 1.8^\circ]$ for $\delta\theta = [30^\circ, 60^\circ, 90^\circ]$ respectively. For this process, all emitters are positioned optimally, $\bm{R}_\mu = (x_0 + ai, aj, 104\text{nm})$. 

Applying the Eq.~(\ref{Mean}) and Eq.~(\ref{Std}) to both the positional and dipole orientation uncertainty results seen in Fig.~\ref{PRL Uncert}d and Fig.~\ref{PRL Uncert}f allows for a direct comparison of the impact of the different sources of experimental noise, as seen in Fig.~\textcolor{blue}{7}, showing the uncertainty in dipole orientation results in a much greater decrease in $\mathcal{G}^{(2)}(0,0)$, whilst other sources of noise produces limited variations.

\begin{figure}[t]
\centering
\includegraphics[width=\textwidth]{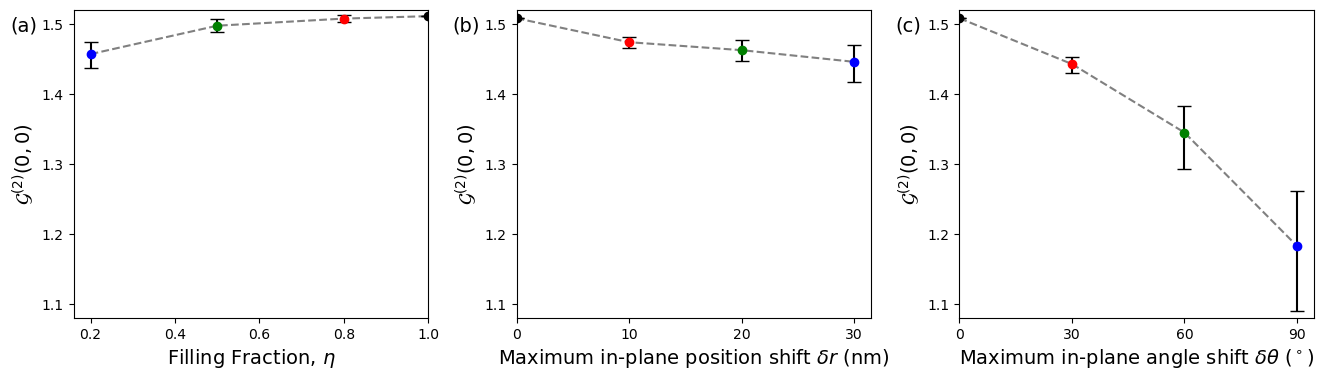}
\caption{Plots showing the mean and standard deviations of \textbf{(a)} Filling fraction, \textbf{(b)} in-plane positional uncertainty, and \textbf{(c)} in plane dipole orientation uncertainty using Eq.~(\ref{Mean}) and Eq.~(\ref{Std}). The error bars here are adjusted to account for the skewness of the distribution using $\sigma_i (1\pm\gamma_i/2)$. The colours in each plot are used to correspond to the distributions seen in Fig.~\ref{PRL Uncert}b, Fig.~\ref{PRL Uncert}d, and Fig.~\ref{PRL Uncert}f.}
\label{Error Bar plot}
\end{figure}

The resulting means ($\mu$), standard deviations ($\sigma$),  and skewness ($\gamma$), as defined in Eq.~(\ref{Stat eqns}), are listed in Table \textcolor{blue}{I} for the different distributions seen in Fig.~\ref{PRL Uncert}.

\begin{table}[h!]
\centering
\begin{tabular}{||c | c | c | c||} 
 \hline
 Experimental Noise & $\mu$ & $\sigma$ & $\gamma$ \\ [0.5ex] 
 \hline\hline
 $\eta = 0.2$ & 1.457 & $1.86\times10^{-2}$ & -0.0742 \\ 
 \hline
 $\eta = 0.5$ & 1.497 & $9.43\times10^{-3}$ & -0.0441 \\
 \hline
 $\eta = 0.8$ & 1.507 & $4.79\times10^{-3}$ & -0.043 \\
 \hline\hline
 $\delta r = 10$nm & 1.474 & $7.65\times10^{-3}$ & -0.137 \\
 \hline
 $\delta r = 20$nm & 1.462 & $1.49\times10^{-2}$ & -0.0911 \\
 \hline
 $\delta r = 30$nm & 1.446 & $2.62\times10^{-2}$ & -0.147 \\
 \hline\hline
 $\delta \theta = 30^\circ$ & 1.442 & $1.11\times10^{-2}$ & -0.270 \\
 \hline
 $\delta \theta = 60^\circ$ & 1.345 & $4.50\times10^{-2}$ & -0.328 \\
 \hline
 $\delta \theta = 90^\circ$ & 1.183 & $8.54\times10^{-2}$ & -0.167 \\
 \hline
\end{tabular}
\label{table:dist values}
\caption{Table summarising the means, standard deviations, and skewnesses for the different sources of experimental noise explored within Fig.~\ref{PRL Uncert}. }

\end{table}

\end{widetext}

\end{document}